\input{aipcheck}

\documentclass[
    ,draft            
  ]
  {aipproc}

\layoutstyle{6x9}

\begin{document}

\title{Long Gamma-Ray Bursts as standard candles}

\classification{98.70.Rz, 98.80.-k, 95.36.+x}
\keywords      {gamma-ray sources; gamma-ray bursts; 
cosmology; dark energy}

\author{Davide Lazzati}{
  address={JILA, University of Colorado, 440 UCB, Boulder 
CO 80309-0440, USA} }

\author{Giancarlo Ghirlanda}{
  address={Osservatorio Astronomico di Brera, via Bianchi 46, 
22807 Merate, LC, Italy}}

\author{Gabriele Ghisellini}{
  address={Osservatorio Astronomico di Brera, via Bianchi 46, 
22807 Merate, LC, Italy}}

\author{Lara Nava}{
  address={Osservatorio Astronomico di Brera, via Bianchi 46, 
22807 Merate, LC, Italy}}

\author{Claudio Firmani}{
  address={Osservatorio Astronomico di Brera, via Bianchi 46, 
22807 Merate, LC, Italy}}

\author{Brian Morsony}{
  address={JILA, University of Colorado, 440 UCB, Boulder 
CO 80309-0440, USA} }

\author{Mitchell C. Begelman}{
  address={JILA, University of Colorado, 440 UCB, Boulder 
CO 80309-0440, USA} }

\begin{abstract}
As soon as it was realized that long GRBs lie at cosmological
distances, attempts have been made to use them as cosmological
probes. Besides their use as lighthouses, a task that presents mainly
the technological challenge of a rapid deep high resolution follow-up,
researchers attempted to find the Holy Grail: a way to create a
standard candle from GRB observables. We discuss here the attempts and
the discovery of the Ghirlanda correlation, to date the best method to
standardize the GRB candle. Together with discussing the promises of
this method, we will underline the open issues, the required
calibrations and how to understand them and keep them under
control. Even though GRB cosmology is a field in its infancy, ongoing
work and studies will clarify soon if and how GRBs will be able to
keep up to the promises.
\end{abstract}

\maketitle


\section{Introduction}

At first glance, gamma-ray bursts (hereafter GRBs), are all but
standard candles. We can quantify this statement by computing the
isotropic equivalent energy, i.e. their energy output in photons
assuming they radiate in every direction with the same properties:
\begin{equation}
E_{\rm{iso}} = \frac{4\pi\,D_L^2}{1+z}{\cal F_{\rm{bol}}}
\end{equation}
where ${\cal F_{\rm{bol}}}$ is the burst bolometric fluence, $z$ its
redshift and $D_L$ its luminosity distance. This quantity is easy to
measure, provided that the burst has been detected and its redshift
measured. A basic knowledge of its spectrum is necessary in order to
perform a bolometric correction, unless a statistical approach is adopted 
\cite{Bloom01}. Compilation of $E_{\rm{iso}}$ for burst samples show 
that this quantity spans, at least, three orders of magnitude
\cite{Frail01,Ghirla04a}. The isotropic equivalent energy is therefore 
all but a good standard candle. Type Ia supernov\ae\ (SNe), as a test
bench, have a RMS scatter of 0.075 decades.

Despite this apparent failure, the interest in GRBs as cosmological
probes did not wane. Another related research branch was the effort to
find a redshift estimate for bursts without optical afterglows. The
real requirement of a cosmological test is to measure the source
distance without measuring its redshift. Such a measure leads to an
estimate of a luminosity distance or angular distance that allows for
comparison with cosmological models.

In this paper I will review the attempts made to use GRBs as standard
candles (\S~2). I will then discuss the Ghirlanda relation (\S~3),
that seem to finally provide us with a candle that is standard enough
to be used to test cosmological models (\S~4). I will finally discuss
future prospects and developments of this idea (\S~5).

\section{Correlations}

\begin{figure}
  \includegraphics[height=.3\textheight]{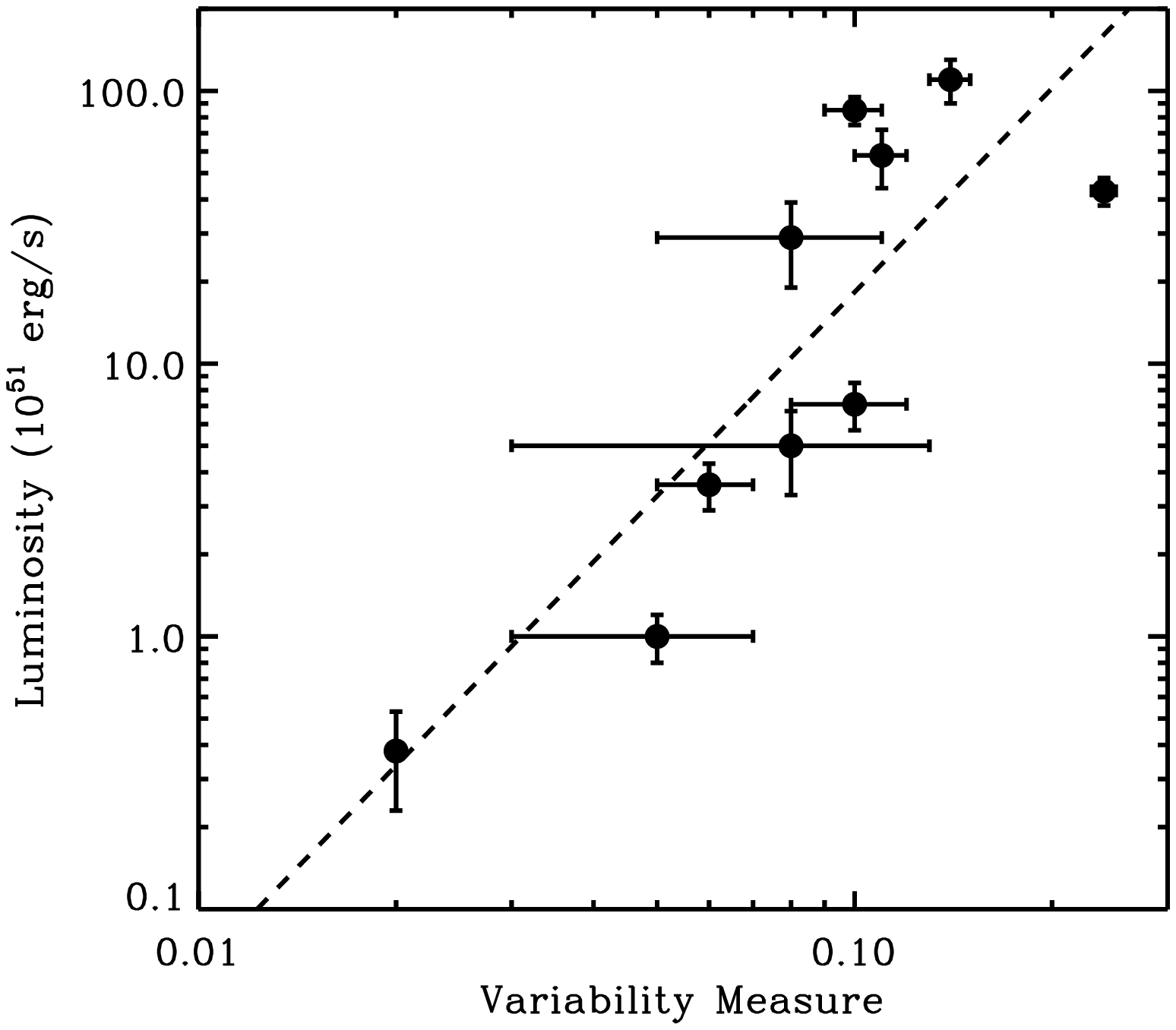}
  \includegraphics[height=.3\textheight]{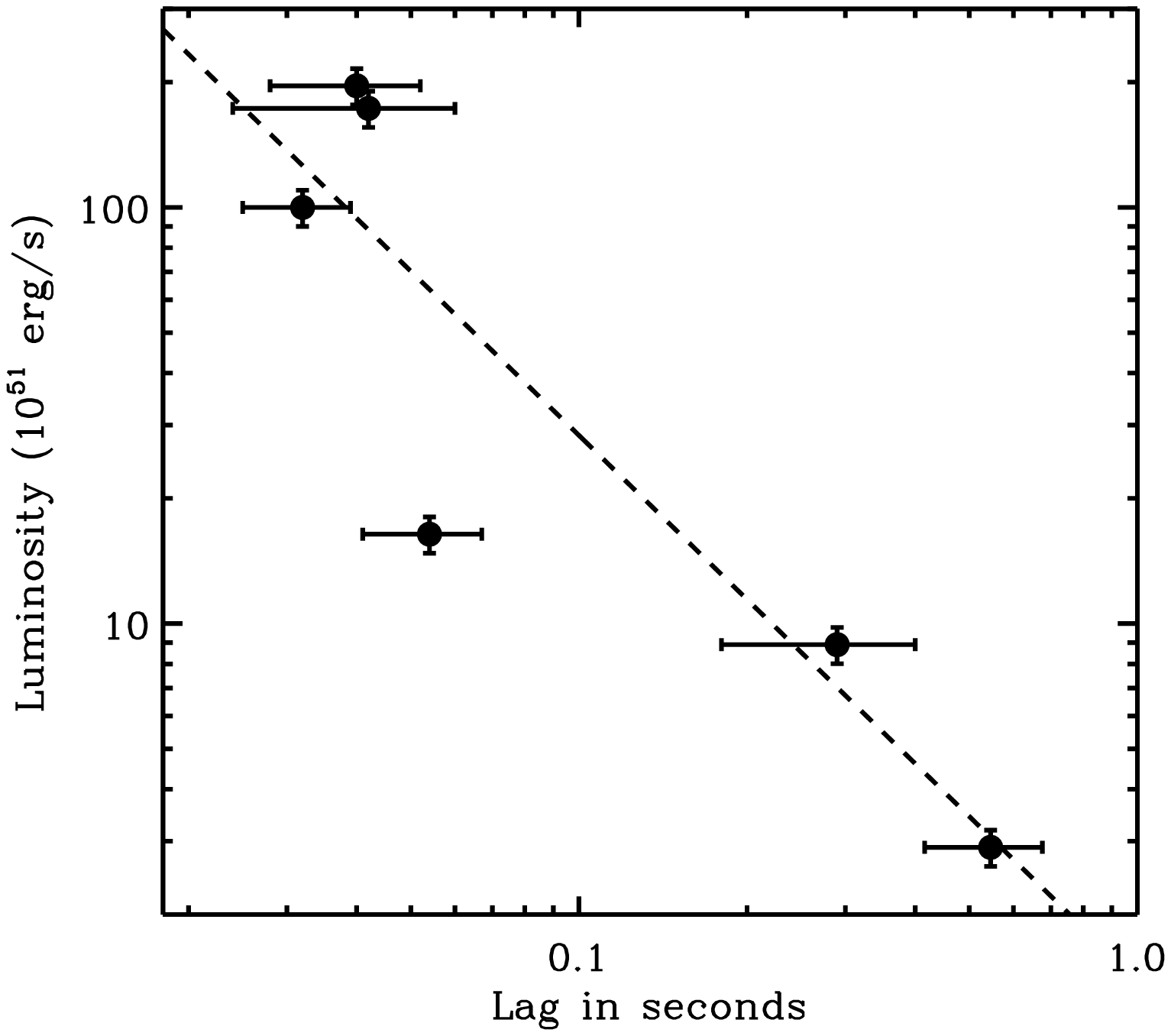} \caption{The
  variability-luminosity (left panel) and lag-luminosity (right panel)
  correlations for a sample of {\it{Beppo}SAX} GRBs. Data taken
  from~\cite{Reichart01,Norris00}.\label{fig:varlum}}
\end{figure}

The first attempt to find a correlation between GRB observables that
could eventually lead to a distance measurement independent of
redshift was the so-called variability-luminosity correlation
\cite{Fenimore00,Reichart01} (see Fig.~\ref{fig:varlum}). It was 
found, for a small sample of {\it{Beppo}}SAX bursts with measured
redshifts, that the intrinsic peak luminosity is correlated with the
amount of variability present in the light curve. Therefore, if the
light curve of a GRB can be observed with enough accuracy, the amount
of variability can be converted into an intrinsic luminosity that,
compared to the flux, yields a distance measurement.

Almost simultaneously, a correlation was discovered also between the
luminosity and the time lag between light curve features at different
frequencies \cite{Norris00} (see Fig.~\ref{fig:varlum}). The time lag
is measured by cross-correlating the light curve of a GRB in two
different energy bands (usually BATSE channels 1 and 3). It is found
that the longer the lag between the two channels, the smaller the peak
luminosity of the event. An attempt to use these correlations for
cosmological purposes yielded poor results, due to the large scatter
of both relations. Only formal limits to the cosmological parameters
could be obtained \cite{Schaefer03}.

A different correlation was discovered after the realization that GRBs
are most likely beamed in a cone rather than isotropic
explosions. Frail et al. \cite{Frail01} and Panaitescu and Kumar
\cite{Panaitescu01} noted independently that the more luminous is the 
GRB, the earlier is its break time (and therefore the smaller its
beaming angle).  If the isotropic equivalent energy $E_{\rm{iso}}$ is
corrected for the beaming angle, a remarkable clustering of the
``true'' energies is obtained. Given the need to know the break time
of the afterglow light curve, it is more difficult to obtain all the
necessary information for a given GRB. Analogously to the other
correlations described above, this correlation is affected by a large
scatter that does not allow its use for cosmological tests
\cite{Bloom03}.

\begin{figure}
  \includegraphics[height=.3\textheight]{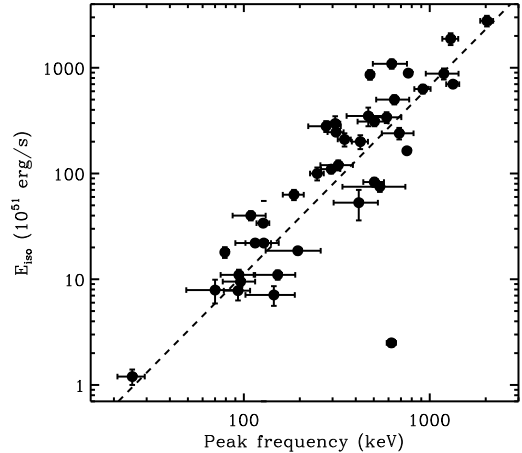} \caption{The Amati
  correlation. Data from \cite{Amati06}.\label{fig:amati}}
\end{figure}

Finally, a new insight into the parameter space was opened by the
systematic analysis of spectra of the prompt emission of
{\it{Beppo}SAX} GRBs \cite{Amati02}. Amati et al. showed that a
remarkable correlation is present between the typical photon frequency
(measured as the peak of the spectrum in $\nu{}F(\nu)$ units) and the
burst isotropic equivalent energy (see Fig.~\ref{fig:amati}). Again,
the scatter of the correlation is too big to allow for any
cosmological use. A debate is open on whether there is a selection
effect increasing the narrowness of the correlation \cite{Nakar05}.

\section{The Ghirlanda Correlation}

\begin{figure}
  \includegraphics[height=.3\textheight]{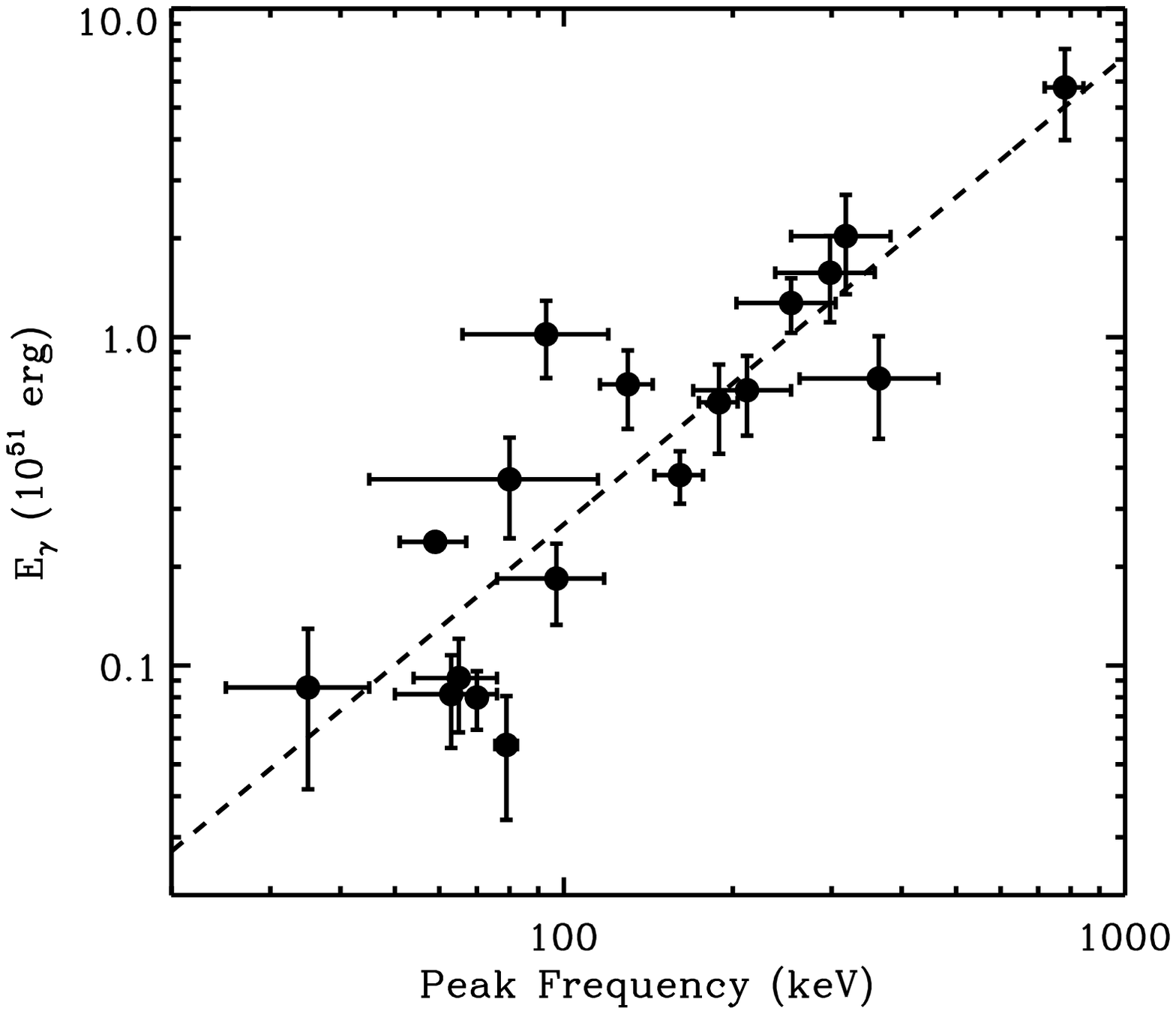}
  \includegraphics[height=.3\textheight]{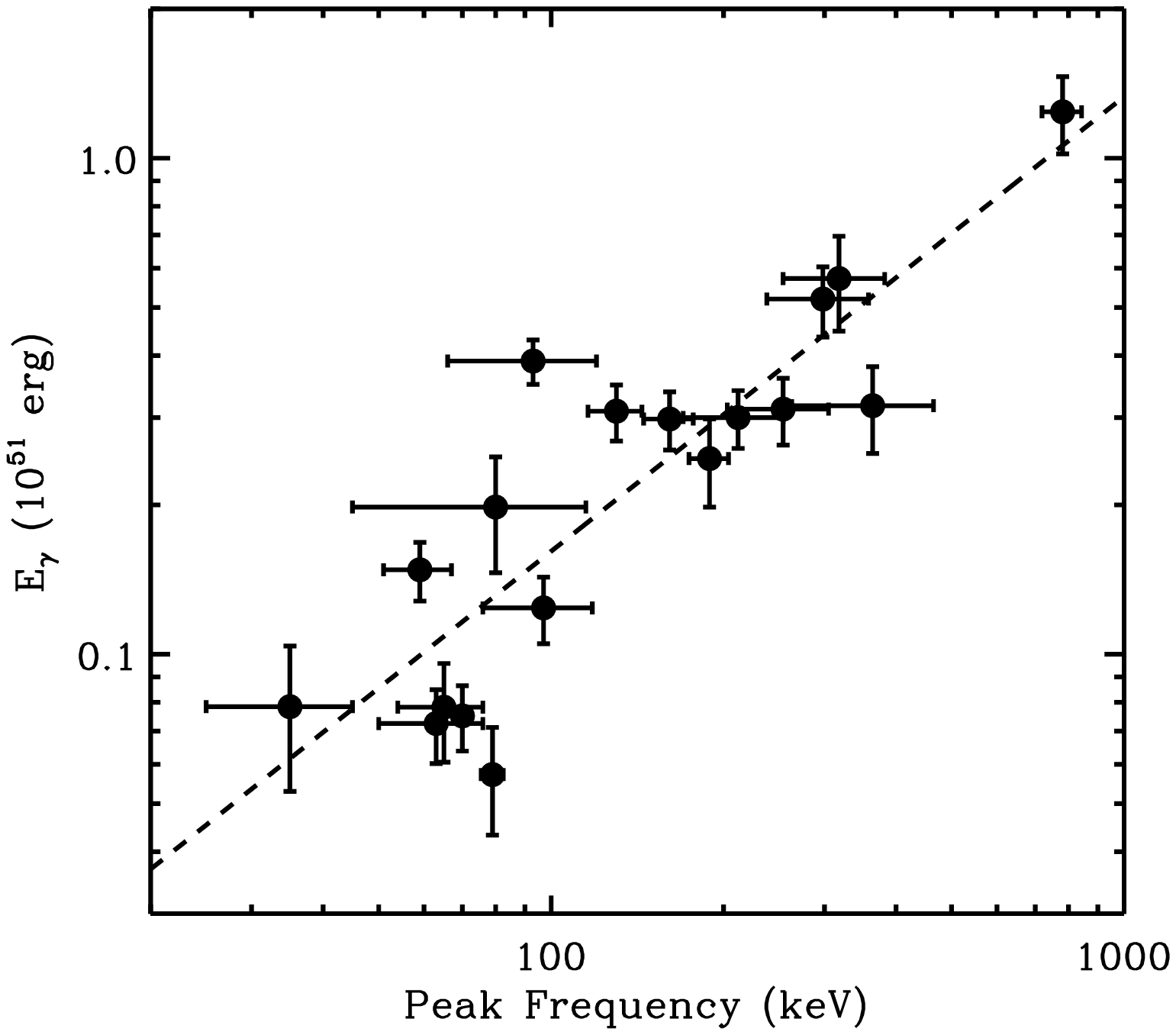} \caption{The
  Ghirlanda correlation. The left panel shows the results assuming a
  uniform environment for all GRBs, while the right panel shows the
  result under the assumption of a wind environment. Data from
  \cite{Nava06}.\label{fig:nava}}
\end{figure}

A dramatic improvement over the previous attempts was achieved by
merging the standard energy results \cite{Frail01,Panaitescu01} with
the spectral results \cite{Amati02}. Ghirlanda and collaborators
\cite{Ghirla04a} noticed that if the Amati plot (Fig.~\ref{fig:amati})
is modified by using the beaming corrected energy rather than the
isotropic equivalent one, a very narrow relation between points is
obtained. The scatter of points around the correlation is impressively
small. A Gaussian fit yielded a $1-\sigma$ dispersion of 0.15 decades,
approximatively twice as large as the scatter of Type Ia SNe. To
counterbalance the small scatter of the correlation, placing a GRB on
the Ghirlanda plot requires a complicated set of observations. The
measure of the typical frequency of photons in the prompt GRB emission
requires broadband coverage, from X-rays to soft $\gamma$-rays. If
such a broad coverage was provided in the past by BATSE and
{\it{Beppo}SAX}, most of the Swift bursts do not have an adequately
large band, unless observed simultaneously by HETE-2. On the other
hand, the measure of the beaming corrected energy requires a knowledge
of the opening angle of the GRB. This can be obtained only with
continuous sampling of the afterglow light curve, possibly in multiple
bands, to unambiguously identify an achromatic break that can be
associated to the geometry of the fireball. Finally, a good measure of
the opening angle of the GRB requires the knowledge of the density
profile of the medium surrounding the GRB. The association of GRBs
with the death of massive stars and simultaneous explosion with Type
Ic SNe \cite{Hjorth03,Stanek03,Malesani04} suggests a stratified
medium surrounding the GRB. On the other hand, external shock fits
generally suggest a flat distribution of density. It can be shown
\cite{Nava06} (see the right panel of Fig.~\ref{fig:nava}) that 
changing the assumption on the environment profile does not destroy
the correlation, but changes its slope. A tighter correlation is found
for a wind environment.

\section{The Ghirlanda correlation as a cosmology tool}

\begin{figure}
  \includegraphics[height=.55\textheight]{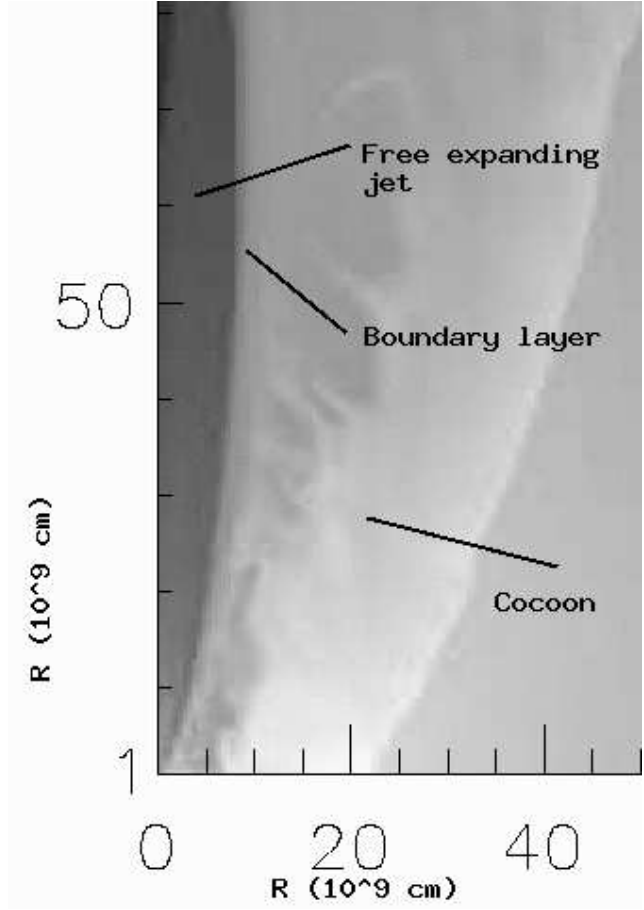} \caption{The
  structure of a relativistic jet expanding inside a massive star
  core.\label{fig:jet}}
\end{figure}

\begin{figure}
  \includegraphics[height=.5\textheight]{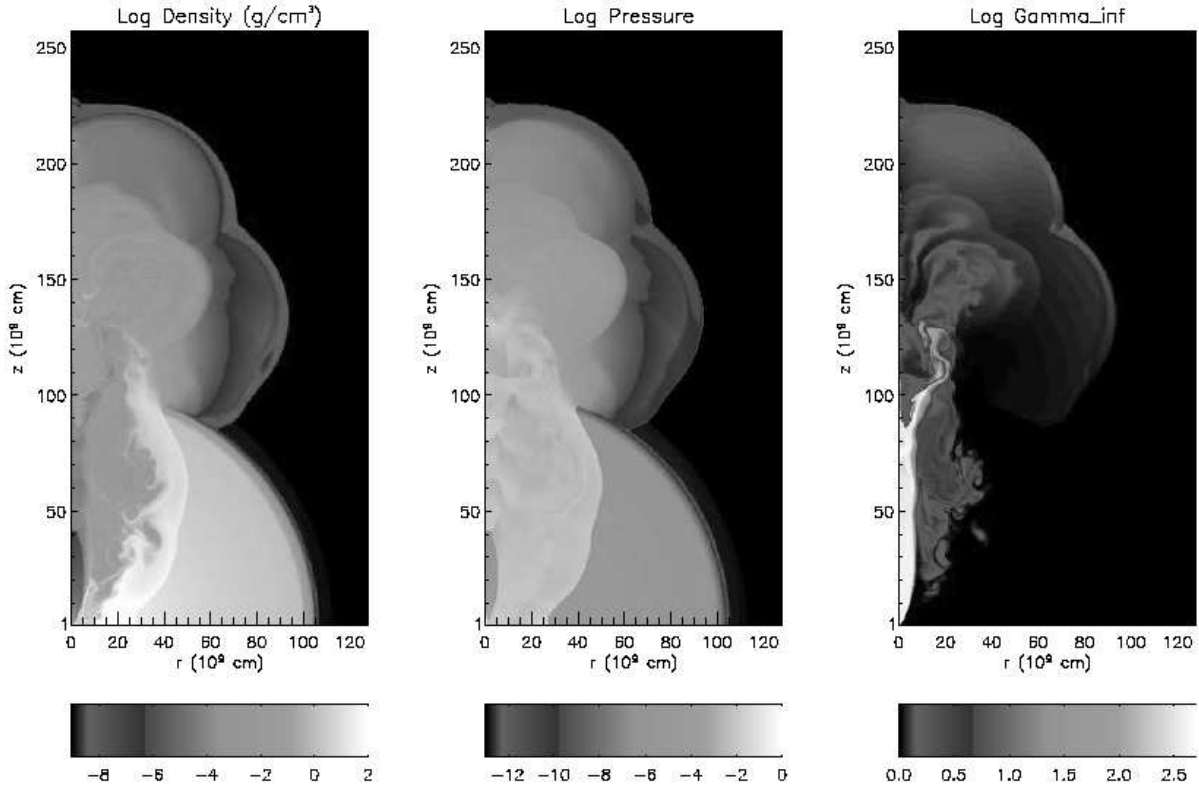} \caption{The
  structure of a relativistic jet as it exits from the surface of the
  progenitor massive star. The left panel shows density, the central
  panel shows pressure and the right panel shows Lorentz factor at
  infinity, or dimensionless entropy. \label{fig:jet2}}
\end{figure}

Thanks to the narrowness of the Ghirlanda correlation, a GRB Hubble
diagram can be produced yielding promising results \cite{Ghirla04b}. A
deep knowledge of the systematic errors is however necessary before a
meaningful constraint on cosmological parameters can be derived
\cite{Friedman05}. Besides the uncertainty on the environment 
stratification and the difficulty in the measurement of the peak
frequency with a relatively narrow band instrument, a major issue is
to understand the relation between the break time in the light curve
and the opening angle of the jet. So far toy models for which the jet
is uniform inside a well-defined opening angle and absent outside have
been used. However, it is expected that this model is not extremely
accurate. If the jet is hydrodynamically confined inside the
progenitor star \cite{Macfadyen99,Lazzati05}, the jet reaching th
surface of the star is expected to be complex (see
Fig.~\ref{fig:jet}). The jet-star interaction creates a high-pressure
cocoon that surrounds and confines the jet. The jet structure evolves
then toward a stable solution with a freely expanding jet core
surrounded by a boundary layer that flows parallel to the jet-cocoon
discontinuity and is in pressure equilibrium with the cocoon. The jet
that emerges on the surface of the star is therefore complex as can be
seen in Fig.~\ref{fig:jet2}, where density, pressure and entropy of
the jet are shown few seconds after the jet breakout (Morsony et
al. in preparation).

\begin{figure}
  \includegraphics[height=.5\textheight]{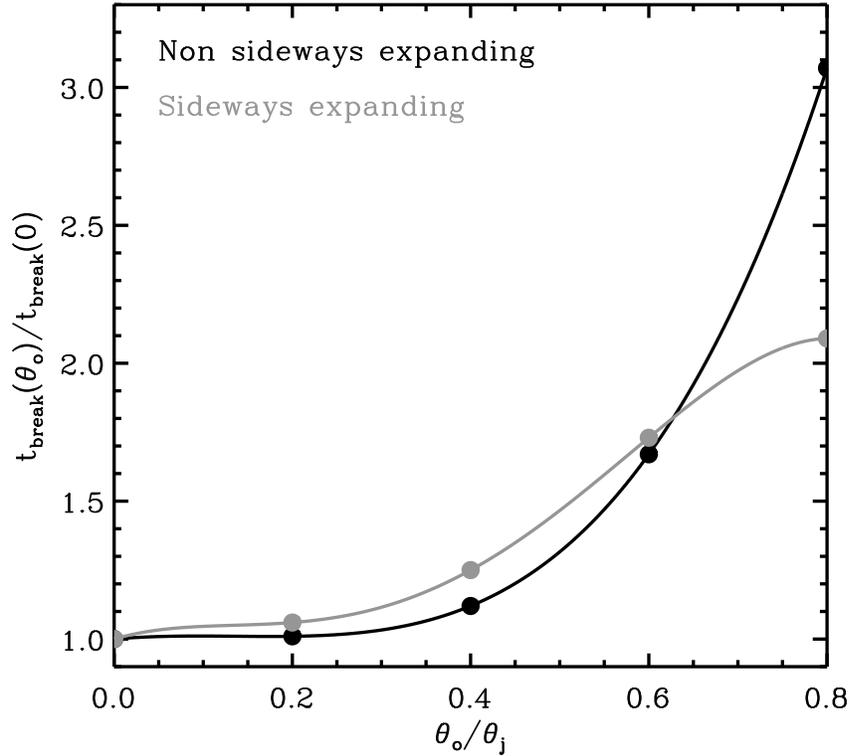} \caption{The dependence
of the break time on the viewing angle of the jet. The black line and
points show the dependence for a non sideways expanding jet while the
gray line and points show the behavior for a sideways expanding jet.
\label{fig:tb}}
\end{figure}

Besides the structure of the jet, another aspect to consider is the
viewing angle of the observer with respect to the jet axis. Rossi et
al.~\cite{Rossi04} showed that, in the simplest assumption of a
conical jet that does not expand sideways, the viewing angle has a
strong influence on the time of the break (Fig.~\ref{fig:tb}). This
uncertainty should propagate and affect the dispersion of the points
around the Ghirlanda correlation. The observed dispersion is too small
and therefore a smaller dependence of the break time from the viewing
angle has to be relevant in the real case. One possibility to avoid
such a large dependence of the break time on the viewing angle is to
assume a so-called structured jet \cite{Rossi04}. In this case, all
the jets are assumed to be alike, but with an energy distribution in
solid angle,
\begin{equation}
\frac{dE}{d\Omega} \propto \theta^{-2} \;\;.
\end{equation}
It can be shown that with such an energy configuration the afterglow
light curve is similar to that of a uniform jet, with an early-time
shallow decay, a late-time fast decay, and a break time that connects
the two regimes. The break time, however, is not related to the
opening angle of the jet but to the viewing angle. As a consequence,
to any break time is associated a single viewing time and a single
value of $E_{\rm{iso}}$. This configuration mimicks a wide low-energy
jet for oservers far from the axis of the beam and mimicks a narrow
powerful jet for observers close to the jet axis. This solution
explains successfully the small scatter of the Ghirlanda
relation. Attempts to unveil the true structure of the jet have so far
been inconclusive. Theoretically, the jet structure may be due to the
turbulent interaction of the jet with the star or to the spreading of
the jet as the star confinement progressively wanes and disappear
\cite{Lazzati05}.

\begin{figure}
  \includegraphics[height=.5\textheight]{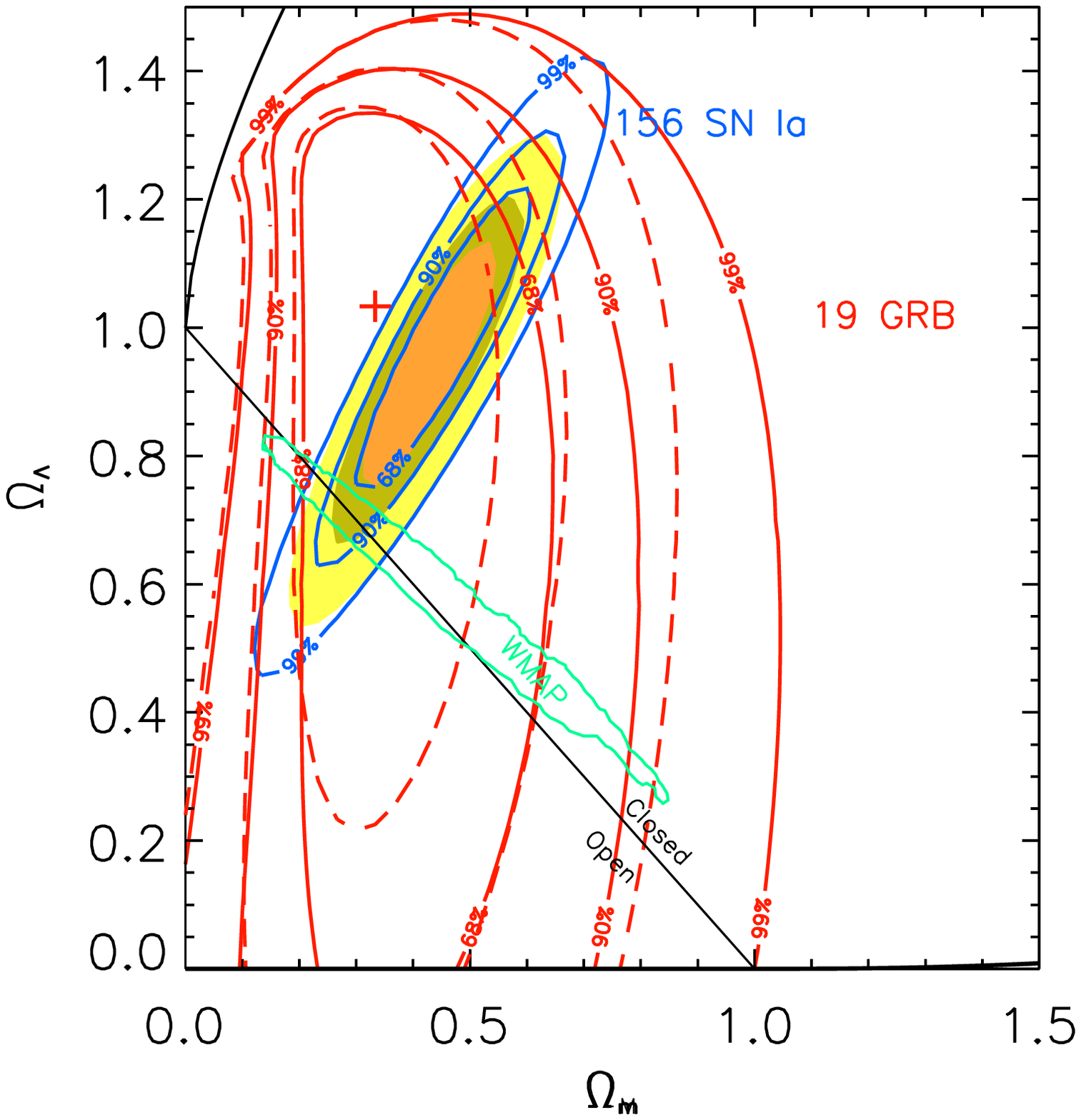}
  \caption{Constraints in the $\Omega_M-\Omega_\Lambda$ plain that can
  be obtained with the application of the Ghirlanda relation to the
  present sample of 19 GRBS (data from \cite{Ghirla06}). The shaded
  contours show the constraints obtained by a simultaneous fit of Ia
  SN data with GRBs.
\label{fig:cosmo}}
\end{figure}

\begin{figure}
  \includegraphics[height=.3\textheight]{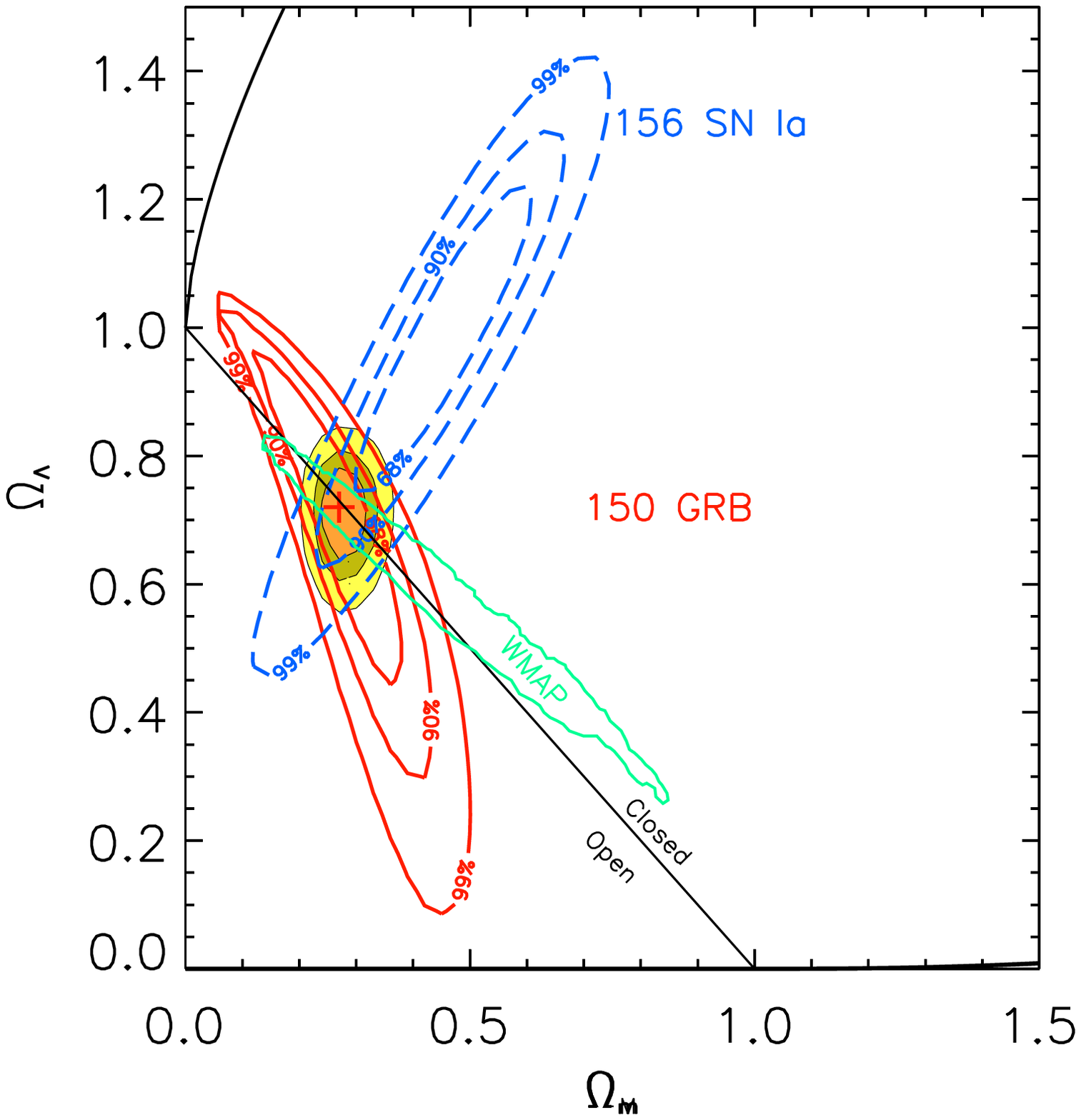} 
  \includegraphics[height=.3\textheight]{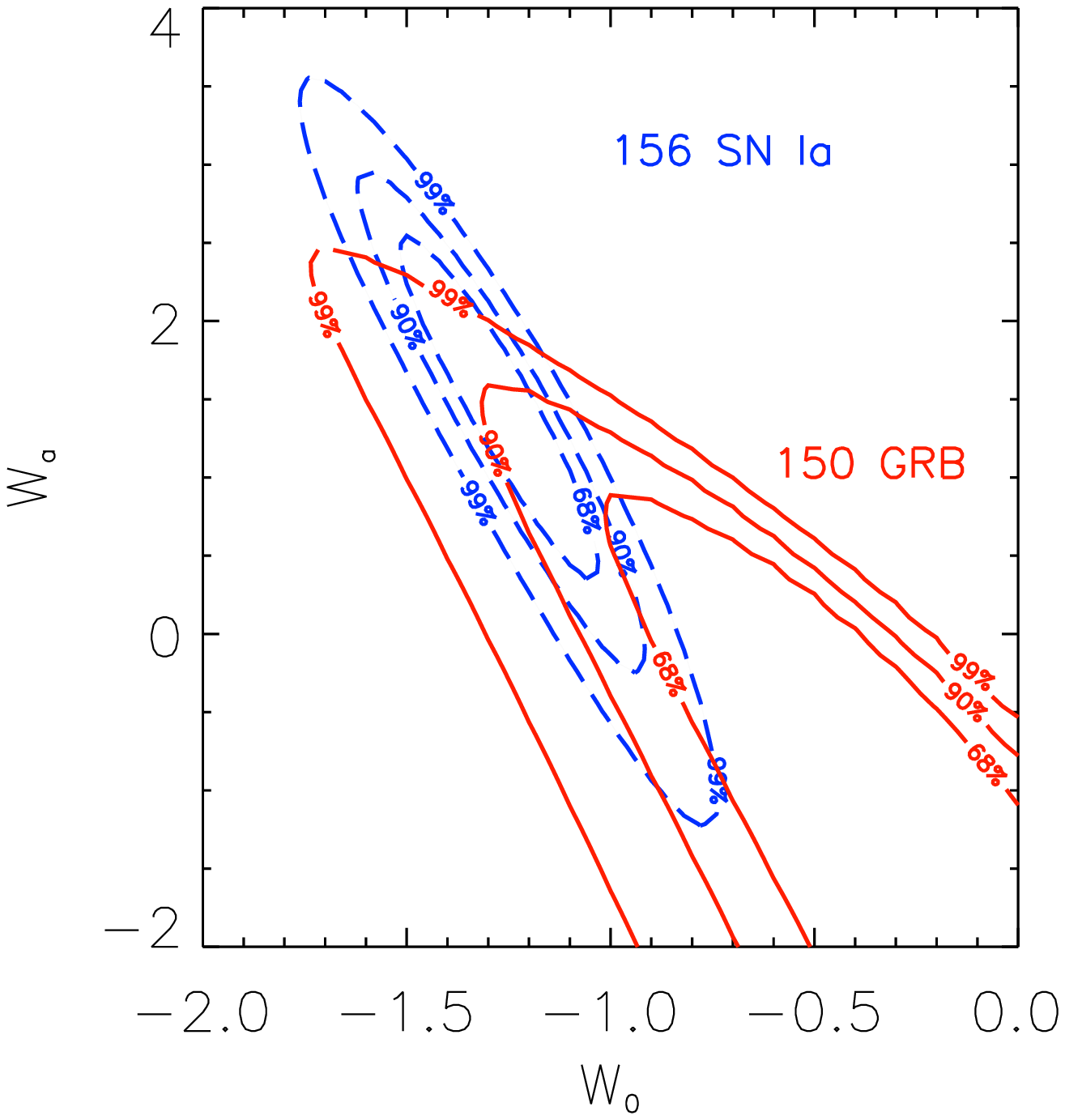} 
\caption{Constraints on the cosmological parameters and on the 
evolution of dark energy from a simulated set of 150 GRBs compared to
analogous constraints obtained from the present Ia SN sample.
\label{fig:cosmo2}}
\end{figure}

Despite these theoretical and observational uncertainties one can
blindly apply the Ghirlanda correlation to derive constraints on the
cosmological parameters
\cite{Ghirla04b,Ghirla06}. Figure~\ref{fig:cosmo} shows the result
that can be derived with the present sample of GRBs. The allowed
region is still wide, mainly due to the paucity of the sample. A
fairer comparison can be obtained by simulating a sample of 150
bursts. Figure~\ref{fig:cosmo2} shows the result of such an
effort. GRBs indeed provide strong constraints on the cosmological
parameters and on the Dark Energy evolution.

\section{Discussion}

The attempt to standardize the GRB candle has a long history. Since it
was realized that GRBs explode at large redshifts, it became of great
interest to understand whether GRBs could be used as distance
indicators. It was soon clear that at first glance GRBs are useless as
distance indicators, since their apparent luminosity spans at least
three orders of magnitude. Soon after, however, it was realized that
correlations between observables exist and that these correlations can
lead to a more standard GRB candle. Attempts with the
variability-luminosity and lag-luminosity correlations yielded
inconclusive results. The large scatter of the Amati relation
hampered, in a similar way, all attempts to use it for cosmological
purposes.

The recent discovery of the Ghirlanda correlation \cite{Ghirla04a}
opened new possibilities. The scatter of points around that
correlation is only approximatively twice as large as that of Type Ia
SNe around the stretching-luminosity correlation. With a suitably
large sample (comparable in number to that of Ia SNe) GRBs could
provide interesting new constraints on cosmological parameters and the
evolution (if any) of the Dark Energy. However, more insight in the
origin of the correlation and its systematic errors has to be obtained
before strong claims can be made. In particular, we still have to
understand clearly the structure of the external medium, the jet
structure and the relation of the jet break time to the opening angle
(or structure) of the jet.

\begin{theacknowledgments}
This work was supported by NSF grant AST-0307502 and NASA
Astrophysical Theory Grant NAG5-12035.
\end{theacknowledgments}



\bibliographystyle{aipprocl} 


\end{document}